\documentclass[prl,singlecolumn,showpacs,preprintnumbers,
amsmath,amssymb]{revtex4}

\usepackage{graphicx}
\usepackage{dcolumn}
\usepackage{bm}

\newcommand{\bfe}{{\bf e}}
\newcommand{\bft}{{\bf t}}

\begin{document}
%
%
%
%
\title  {DUAL SUPERCONDUCTORS AND SU(2) YANG-MILLS }

\author{Antti J. Niemi } 

\email{niemi@teorfys.uu.se}

\affiliation{ Department of Theoretical Physics, 
Uppsala University, P.O. Box 803, S-75108, Uppsala, Sweden }

\homepage{http://www.teorfys.uu.se/people/antti}

\date{\today}

\begin{abstract}
We propose that the SU(2) Yang-Mills theory can be interpreted 
as a two-band dual superconductor with an interband Josephson 
coupling. We discuss various consequences of this interpretation 
including electric flux quantization, confinement of vortices
with fractional flux, and the possibility that a closed vortex
loop exhibits exotic exchange statistics.
\end{abstract}


\maketitle

%
%

Color confinement remains one of the great mysteries of modern
high energy physics \cite{clay}. Several approaches have been 
proposed to explain why and how a SU(N) Yang-Mills 
theory supports confining colored strings, including a variety
of dual superconductor and central vortex models \cite{conf}. But until 
now the relevant dynamical string degrees of freedom 
have not been identified. 

Recently it has been proposed \cite{fad1} that the string 
variables in a Yang-Mills theory could relate to
an appropriate decomposion of the gauge field.
In the simplest case of SU(2) considered here,
we start by separating $A^a_\mu$ ($a=1,2,3$) into 
its U(1) Cartan component $A^3_\mu \equiv A_\mu $ while the off-diagonal 
$A^i_\mu$ ($i=1,2$) we combine into the 
complex variable 
\[
W^{\pm}_\mu = A^1_\mu \pm i A^2_\mu
\]
We introduce the U(1) covariant derivatives 
\[
D^{ij}_\mu = \delta^{ij}
\partial_\mu - g \epsilon^{ij}A_\mu
\ \ \ \ \ {\rm and} \ \ \ \ \ 
D^\pm_{\mu } = \partial_\mu \pm i g A_\mu
\]
and with $F_{\mu\nu}
= \partial_\mu A_\nu - \partial_\nu A_\mu$ we expand 
the Yang-Mills Lagrangian as follows,
\begin{eqnarray}
&& - \frac{1}{4}(F^a_{\mu\nu})^2 + 
\frac{\xi}{2} (D^{ij}_\mu A^j_\mu)^2 +  {\cal L}_{FP}
= - \frac{1}{4}(F^3_{\mu\nu})^2 
- \frac{1}{4} F^+_{\mu\nu} F^-_{\mu\nu} +
\frac{\xi}{2} (D^{ij}_\mu A^j_\mu)^2 + {\cal L}_{FP}
\nonumber
\\
= && -\frac{1}{4}( F_{\mu\nu} + G_{\mu\nu})^2 
- \frac{1}{2} F_{\mu\nu}G_{\mu\nu} + \frac{1}{2}
|D^+_\mu W_\nu|^2
- \frac{1}{2} |D^+_\mu W_\mu|^2 + 
\frac{\xi}{2} |D^+_\mu W_\mu|^2 
+ {\cal P}^\star D_+^2 \eta 
+ {\cal P} D_-^2 \eta^\star 
\nonumber
\\
&& + \frac{g^2}{2} [{\cal P}^\star W_\mu^+ W_\mu^+ 
\eta + {\cal P} W_\mu^+ W_\mu^-
\eta^\star - {\cal P}^\star W^+_\mu W^+_\mu \eta^\star 
- {\cal P} W^-_\mu W^-_\mu \eta ]
\label{act1}
\end{eqnarray}
Here we have explicitely displayed a gauge fixing
and Faddeev-Popov ghost (${\cal P}, \eta$) \cite{pac}
{\it only} for the off-diagonal 
$A_\mu^{1,2}$ components. The remaining Cartan U(1) gauge 
invariance can be fixed similarly.  But we shall
find that this U(1) invariance can also be eliminated 
directly, by casting the Lagrangian in a form that involves 
only manifestly U(1) invariant variables. In the following 
we shall mostly exclude the ghosts from explicit equations. 
Their inclusion has very little if any conceptual impact.
 
For the present purposes it is important to note that the 
sole effect of the fourth term in the {\it r.h.s.}
of (\ref{act1}) is a redefinition of the gauge parameter 
$\xi \to \xi-1$ in the gauge fixing (fifth) term. Thus 
these terms can be combined. In particular, in the 
$\xi \to \infty$ limit the fourth term
becomes entirely immaterial. This limit yields the 
Landau version of the maximal abelian gauge 
\[
(\partial_\mu + ig A_\mu) W_\mu = 0
\] 
which we shall assume has been imposed. We note that this 
gauge condition appears
as the variational equation when one locates the extrema of 
the functional
\begin{eqnarray}
\int \rho^2 = \int ( \rho_1^2 + \rho_2^2) 
= \int [(A_\mu^1)^2 + (A_\mu^2)^2]
= \int W_\mu W_\mu^*
\label{gf}
\end{eqnarray}
with respect to the full SU(2) gauge transformations. Thus, even 
though the field $\rho$ is in general a gauge dependent quantity,
its extremum values with respect to gauge transformations are 
manifestly gauge independent when the theory is subject to the
Landau version of maximal abelian gauge. Recent numerical investigations 
indicate that in strong coupling regime the extrema
of $\rho$ are {\it nonvanishing} \cite{val}

The antisymmetric tensor 
\[
G_{\mu\nu} = (ig/2) ( 
W_\mu W^\star_\nu - W_\nu W^\star_\mu)
\]
modifies the Cartan part of the Lagrangian according to
\begin{equation}
F_{\mu\nu}^2 \ \to \ (F^3_{\mu\nu})^2 = (F_{\mu\nu} + G_{\mu\nu})^2 +
2 G_{\mu\nu} F_{\mu\nu}  
\label{dirac}
\end{equation}
Dirac proposed in \cite{dirac} that this is in essence
the way how Maxwellian electrodynamics should be improved to accout
for both electric and magnetic sources, he identified 
$G_{\mu\nu}$ as a string tensor, the source term 
for the magnetic four-current. Here we reconcile ourselves 
with the customary terminology to describe confinement
in magnetic dual variables \cite{conf}. 
Consequently we interpret $F_{\mu\nu}$ 
as a closed {\it dual} field strength tensor, and 
view 
\[
E_i = - G_{oi} \ \ \ \ \ \ {\rm and} \ \ \ \ \ \  
H_i = - 1/2 \epsilon_{ijk} 
G_{jk}
\] 
as the electric and 
magnetic components of the (non-closed electric) string tensor
$G_{\mu\nu}$, in a dual version of Dirac's electrodynamics.

We now proceed to the third term on
the {\it r.h.s.} of (\ref{act1}), the sole remaining
term with no ghost contributions.  We 
follow \cite{fad1} and introduce a complex vector 
field $\bfe_\mu$ with 
\[
\bfe_\mu \bfe_\mu  =  0 \ \ \ \ \ \ {\rm and}
\ \ \ \ \ \ \bfe_\mu \bfe^*_\mu = 1
\]
We also introduce two complex scalar fields $\psi_1$ and $\psi_2$,
and decompose the off-diagonal $W_\mu = A_\mu^1 + i A_\mu^2$ into
\cite{fad1}
\begin{equation}
W_\mu \ = \ \psi_1 \bfe_\mu + \psi^\star_2 \bfe^\star_\mu
\label{decom}
\end{equation}
When we substitute this in the third term of (\ref{act1})
we get
\begin{equation}
|(\partial_\mu + ig A^3_\mu) W_\nu|^2 = 
|(D^+_\mu + i \Gamma_\mu) \psi_1|^2 
+ |(D^-_\mu + i \Gamma_\mu) \psi_2|^2
+ \rho^2 |(\partial_\mu - i \Gamma_\mu) \bfe_\nu|^2 + 
\psi_1 \psi_2 (\partial_\mu \bfe_\nu)^2 + \psi_1^\star
\psi_2^\star ( \partial_\mu \bfe^\star_\nu)^2
\label{supc}
\end{equation}
where $\rho$ is defined by (\ref{gf}) and
\[
\Gamma_\mu = i \bfe_\nu \partial_\mu \bfe^*_\nu
\] 
is a (composite) gauge field for the
internal U(1) rotation 
\[
\psi_{1,2} \to e^{i\alpha} \psi_{1,2} \ \ \ \ \ \ {\rm and}
\ \ \ \ \ \ \bfe_\mu \to e^{-i\alpha} \bfe_\mu
\] 
which leaves the decomposition
(\ref{decom}) intact \cite{fad1}.

Here it is interesting to also spell out the four last
ghost interaction terms on the {\it r.h.s.} of (\ref{act1}). 
We define the ghost number operator 
\[
\hat N = {\cal P}^\star \eta + {\cal P} \eta^\star
\]
and
combine these terms into
\begin{equation}
\frac{g^2}{2} \rho^2 {\hat N} - g^2 \psi_1 \psi_2^\star
{\cal P}^\star \eta^\star - g^2 \psi_1^\star \psi_2 {\cal P} \eta
\label{ghost1}
\end{equation}
(Note that $\rho$ gives an effective mass for the ghosts.)
The {\it r.h.s.} of (\ref{supc}) and (\ref{ghost1}) admit
a direct interpretation
in terms of the Landau-Ginzburg free energy of a two-gap superconductor, 
employed to describe high-$T_c$ materials \cite{egor1}, \cite{egor2}
such as the widely studied {\tt MgB$_2$} \cite{htc}.  Indeed, 
the first two terms 
in the {\it r.h.s.} of (\ref{supc}) are conventional
kinetic terms for two oppositely charged
Cooper pairs, except that now these Cooper pairs are also minimally 
coupled to the additional U(1) gauge field $\Gamma_\mu$ 
so that we have a local $\rm U_A(1) \times U_\Gamma (1)$ symmetry. 
This additional gauge field can be viewed as a Josephson current
which originates in a gauged nonlinear 
sigma-model, in the present case with a O(3,1) structure, which is  
defined by the third term in the {\it r.h.s.} of (\ref{supc}). 
In a condensed matter context such a term could 
model fluctuations in the underlying material
lattice structure. Finally, the last two terms on the 
{\it r.h.s.} of (\ref{supc}) together with the last two terms
in (\ref{ghost1}) have the standard form of interband Josephson 
transition terms between the two Cooper pair condensates  
\cite{legg}. When we group the condensates into multiplets
in two alternative ways, with the $\rm U_A(1)$ charge of $(\psi_1 , 
\psi_2^\star)$ opposite to that of $(\psi_1^\star , \psi_2)$,
and the $\rm U_\Gamma (1)$ charge of $(\psi_1 , \psi_2)$
opposite to that of $(\psi_1^\star , \psi_2^\star)$, we find
that the last two terms in (\ref{supc}) are interband Josephson
transition terms between the $\rm U_A(1)$ multiplets, while the 
last two terms in (\ref{ghost1}) describe interband transitions
between the $\rm U_\Gamma(1)$ multiplets. 

We are particularly interested in
vortex-like configurations in (\ref{dirac}), 
(\ref{supc}), since these are
the natural candidates for describing confining strings.
For this we define 
\[
\psi_i = \rho_i \exp(i 
\theta_i) \ \ \ \ \ \ (i=1,2)
\] 
where according to (\ref{decom}),
the $\rho_{1,2}$ are the same fields as those in (\ref{gf}). 
We introduce the supercurrent
\begin{eqnarray}
J_\mu = && g A_\mu + \frac{i}{2(\rho^2 + \hat N)} ( W_\nu \partial_\mu
W_\nu^\star - \partial_\mu W_\nu W_\nu^\star 
- {\cal P}^\star
\partial_\mu \eta + \partial_\mu {\cal P}^\star \eta + {\cal P}
\partial_\mu \eta^\star - \partial_\mu {\cal P} \eta^\star
)
\nonumber
\\
&& \rightarrow  \ gA_\mu + \frac{\rho_1^2}{\rho^2} (\partial_\mu \theta_1
+ \Gamma_\mu) - \frac{\rho_2^2}{\rho^2} (\partial_\mu \theta_2
+ \Gamma_\mu)
\label{sc1}
\end{eqnarray}
where in the last step we have removed all ghost dependence, since 
the only role of ghosts is to BRST-improve the current.
We note that $J_\mu$ is gauge invariant both under 
$\rm U_A(1)$ and $\rm U_\Gamma (1)$. 
We also introduce the 
three component unit vector 
\begin{equation}
\bft = \frac{1}{\rho^2} ( \psi_1^\star \ \psi_2)
\vec \sigma \left( \begin{array}{c} \psi_1 \\ 
\psi_2^\star \end{array} \right) =
\frac{1}{\rho^2} \left( \begin{array}{c}
2 \rho_1 \rho_2 \cos(\theta_1 + \theta_2)  \\ -2 \rho_1 \rho_2
\sin(\theta_1 + \theta_2) \\ \rho_1^2 - \rho_2^2 \end{array} \right)
\label{deft}
\end{equation}
which is invariant under the Cartan $U_A(1)$. By defining 
\[
(\nabla_\mu \bft)_k = (\delta_{kl}\partial_\mu - 2 \epsilon_{kl3}
\Gamma_\mu) t_l 
\]
we can then represent (the $\xi \to \infty$ limit of) the entire SU(2)
Yang-Mills Lagrangian (\ref{act1}) in terms of Cartan $U_A(1)$ invariant 
variables \cite{egor1} (we do now write ghosts explicitely): 
For the first two terms in 
the {\it r.h.s.} of (\ref{act1}) which correspond to Dirac's extension 
of (dual) electromagnetism we get
\begin{equation}
(F_{\mu\nu} + G_{\mu\nu})^2 + 2 F_{\mu\nu} G_{\mu\nu} =
H_{\mu\nu}^2 + 4i \rho^2 t_3 (\bfe_\mu \bfe_\nu^\star -
\bfe_\nu \bfe^\star_\mu) H_{\mu\nu} + \rho^4 t_3^2
\label{giact1}
\end{equation}
\vskip -0.6cm
\begin{equation}
H_{\mu\nu} = \partial_\mu J_\nu - \partial_\nu J_\mu + \frac{1}{2}
t_3 ( \partial_\mu \Gamma_\nu - \partial_\nu \Gamma_\mu) + \frac{1}{4}
\epsilon_{abc} t_a \nabla_\mu t_b \nabla_\nu t_c
\label{H}
\end{equation}
and the third term in the {\it r.h.s.} of (\ref{act1}) gives; see
(\ref{supc})
\begin{equation}
|(\partial_\mu + igA^3_\mu)W_\nu|^2 =
(\partial_\mu \rho)^2 + \frac{1}{4} (\nabla_\mu \bft)^2 
+ 4\rho^2 J_\mu^2
+ \rho^2 |(\partial_\mu - i\Gamma_\mu) \bfe_\nu|^2 +
\frac{\rho^2}{2} [ t_+ (\partial_\mu \bfe_\nu)^2
+ t_- (\partial_\mu \bfe^\star_\nu)^2 ]
\label{giact2}
\end{equation}
where $t_{\pm} = t_1 \pm i t_2$.

In the case of superconducting materials one often implements
the London limit, which assumes that $\rho_1$ ad $\rho_2$ are 
non-vanishing constants. In a superconductor they 
describe the densities of
the Cooper pairs, and under appropriate conditions 
the London limit is a reasonable approximation.
But in the present case these fields
relate to extrema of (\ref{gf}) under gauge transformations, and the
SU(2) gauge invariance of the London limit must be justified separately, 
for example by numerical investigations \cite{val}. Consequently
here the London limit is solely an
analytically tractable simplification of (\ref{giact1}), (\ref{giact2}).

For asymptotically constant and non-vanishing densities $\rho_1$
and $\rho_2$, the abelian two-condensate Higgs model supports 
line vortices \cite{lines}. We wish to call them {\it egorons}, and
we {\it assume} that they are not removed by our additional terms.
These egorons are then natural candidates
for describing confining strings, and we now analyze their properties:

We compute the (electric) flux of an egoron
as follows: In the London limit the supercurrent $J_\mu$ describes
a massive degree of freedom, which means that we have a Meissner
effect with inverse London penetration length $\propto \rho$. We
integrate $J_\mu$ along a closed path which encircles 
the egoron at a distance much larger than the London penetration 
length. On this path $J_\mu$ then vanishes, and we 
get from (\ref{sc1}) for the flux
\begin{equation}
\Phi = \oint A = \frac{\rho_2^2}{g\rho^2} \oint \partial_\mu
\theta_2 - \frac{\rho_1^2}{g\rho^2} \oint \partial_\mu
\theta_1 \ + \ \frac{\rho_2^2 - \rho_1^2}{g\rho^2}  \oint \Gamma
= \frac{2\pi}{g \rho^2} ( \rho_2^2 N_2 - \rho_1^2 N_1 )
-  \frac{t_3}{g}\oint \Gamma
\label{flux1}
\end{equation}
where $\Delta \theta_{1,2} = 2\pi N_{1,2}$ (with $N_{1,2}$ integers)
are the angular increments in the phases of $\psi_{1,2}$
when we encircle the ($N_1,N_2$) egoron once.

Notice that in the absence of the Josephson contribution
the flux appears to have arbitrary values, depending on the
relative values of the asymptotic densities $\rho_1$ and $\rho_2$
\cite{egor2}. This is in a stark contrast with the 
quantization of magnetic flux of the Abrikosov vortex in the
single-component abelian Higgs model. In order
to further study the flux quantization properties 
of egorons we rewrite the first two terms in the {\it r.h.s.} 
of (\ref{supc}) as
\begin{equation}
|(D^+_\mu + i \Gamma_\mu) \psi_1|^2 
+ |(D^-_\mu + i \Gamma_\mu) \psi_2|^2
= (\partial_\mu \rho_1)^2 + (\partial_\mu \rho_2)^2
+ \rho_1^2 (g A_\mu + \Gamma_\mu + \partial_\mu \theta_1)^2 
+ \rho^2(gA_\mu - \Gamma_\mu - \partial_\mu \theta_2)^2
\label{sc2}
\end{equation}
In the London limit with both $\rho_1$ and $\rho_2$ non-vanishing,
the two supercurrents in (\ref{sc2})
\begin{equation}
j^{1,2}_\mu = gA_\mu \pm ( \Gamma_\mu + \partial_\mu \theta_{1,2})
\label{scura}
\end{equation}
do not describe independent field degrees of freedom,
but are nevertheless massive composites that exhibit the
Meissner effect. Consequently we can compute the 
(electric) flux of a linear ($N_1,N_2$) egoron by 
integrating both (\ref{scura}) over a closed path 
which encircles the egoron at a distance much larger 
than the inverse London penetration lengths 
$\sim \rho_{1,2}$. This gives
\begin{equation}
\Phi = \oint A = \frac{\pi}{g} (N_2 - N_1) \ \ \ \ \& \ \ \ \
\oint \Gamma = - \pi (N_1 + N_2)
\label{flux2}
\end{equation}
For consistency we note that when we substitute this 
in (\ref{flux1}) we recover (\ref{flux2}). 

(A similar computation in the ghost sector with nonvanishing ghost number
also leads to $\Phi$ in (\ref{flux2}), but with $N_1$ and $N_2$ now 
referring to the 
phase increments in the ghost fields.)

The result (\ref{flux2})
reveals that in the presence of an interband Josephson flow 
the flux of an egoron becomes quantized. 
However, for the simplest egorons with $(N_1,N_2) = (\pm 1,0) 
$ or $(0, \pm 1)$ we still obtain a flux quantization
in {\it half} units of an Abrikosov vortex. But 
for an egoron with finite energy per unit length there is also
an interband Josephson flow carried 
by $\Gamma_\mu$. Whenever this interband flow 
vanishes ($\Gamma_\mu = 0$), for finite energy per unit 
length we must have $N_1 = - N_2 = - N$ so that the (dual) flux 
$\Phi$ of an egoron acquires a quantization in integer 
units of the Abrikosov
vortex, $\Phi = 2\pi/g N$. 

We observe that in the absence of any interband Josephson 
coupling the present considerations lead to a curious 
(algebraic) confinement 
of egorons: If we set $\Gamma_\mu \equiv 0$ we conclude from
(\ref{sc2}) that an isolated $(0,\pm 1)$ or $(\pm 1,0)$ 
egoron has a logarithmically divergent energy
per unit length, while (\ref{flux1}) implies that its (electric) 
flux is {\it a priori} an arbitrary fraction of the Abrikosov 
flux. But a neutral bound state egoron of the form 
$(\pm 1 , \mp 1)$ has 
a finite energy, and its flux is quantized in integer units of 
the Abrikosov vortex.

We now wish to bend a finite length linear egoron into a toroidal
ring. But before joining the ends we twist the egoron once around 
its core. The closed ring then acquires a non-trivial 
self-linkage, and numerical simulations indicate that such knotted 
egorons can be stable, at least if we account 
for a definite, possibly radiatively induced correction 
term \cite{sami}. The structure 
of a closed knotted egoron is best described in terms of the unit vector 
$\bft$ in (\ref{deft}). For a static finite energy egoron
it defines a mapping from the compactified $R^3 \sim S^3 \to S^2$
and $\pi_3(S^2) \sim Z$ computes the egorons self-linking number; 
We refer to \cite{nature}, \cite{sami} for details. 
The asymptotic large distance vacuum value of $\bft$ is 
determined by minimizing the Landau-Ginzburg free energy, 
for example the last term of (\ref{giact1}) in soliture 
would identify the asymptotic value of $\bft$ to be the
circle $t_3 = 0$ corresponding to equal asymptotic densities
$\rho_1 = \rho_2$. But in general $t_3 \not= \pm 1$, 
as we only expect that both asymptotic 
densities $\rho_1$ and $\rho_2$ are non-vanishing \cite{val}.

A fixed asymptotic value $t_3\not= \pm1$ defines a circle on the target
$S^2$ and leaves us with a global U(1) invariance in
the phase of $t_{\pm}$. In a confining theory there are no
massless modes, and consequently this global invariance must 
become broken. This is achieved by the interband Josephson 
couplings in (\ref{supc}), (\ref{ghost1}), or alternatively we 
simply combine the phase of $t_\pm$ with $\Gamma_\mu$ to form 
a massive composite with $\bfe_\mu$.

A massive $t_\pm$ identifies the vacuum with a definite
point $\bft_0 \in S^2$, which we assume is different from the poles 
at $t_3 = \pm 1$. The center of the egoron corresponds
to the preimage of the antipodal $- \bft_0$. This implies
that the egoron can be viewed as a double-stranded composite of
two mutually linked closed $(N_1,0)$ and $(0,N_2)$ constituent
egorons in the 
fields $\psi_1$ and $\psi_2$, respectively. The core of 
$(N_1,0)$ is the preimage of the north-pole $t_3=+1$
where $\rho_2$ vanishes, and the core of $(0,N_2)$ is 
the preimage of the south-pole $t_3 = -1$ where $\rho_1=0$.
The relative linking number of these two closed constituent
egorons coincides with the self-linking number of the 
knotted composite egoron. This double-stranded structure of 
egorons has also been observed in
numerical simulations \cite{lines}; see \cite{eba} for a detailed
analysis.

The two closed $(N_1,0)$ and $(0,N_2)$ constituent egorons can be
pulled through each other, provided we do not cross or 
otherwise alter their relative linking number: They can not 
break and rejoin, become split, or cross each other without the 
self-linkage structure of the composite egoron becoming destructed.
This non-destructibility is direct consequence of 
the fact that their cores are the embeddings of two different $S^1 \in
R^3$, corresponding to the preimages of $t_3=\pm 1$, respectively.
We now argue that as a consequence 
the exchange statistics of the two
constituent egorons relates to the exchange 
statistics of anyons on a plane \cite{lerda}. Indeed,
the exchange 
of $N$ pointlike plane particles which leads to Artin's 
braid group of $N$ objects, is often pictured in terms
of vertical strands that describe the time evolution of the particles,
with each strand corresponding to the world-line of a particle \cite{lerda}.
Artin's braid group is generated by elementary moves that braid
these strands around each other in a definite manner. This pictorial
representation relates directly to the linking of 
(non-destructible) loops in three dimensions. For this, simply
draw the strands as lines on the surface of a cylinder
and identify the initial and final points of the individual
strands. The world-lines of the particles now become closed loops on
the surface of the cylinder, and the nontrivial braiding translates
to the nontrivial linking of the loops. But such loops
are special cases of general closed loops embedded in $R^3$, now 
wrapped around the surface of the cylinder. This suggests that more generally,
there must be a relation between the braiding of world-lines and 
the linking of closed loops in $R^3$. Indeed, one can show 
that Artin's braid group appears as a subgroup of motions 
when we exchange (non-destructible) three dimensional closed 
loops \cite{viro}. It is generated by a motion
which corresponds to the pulling of two constituent egorons
through each other without crossing or otherwise changing their relative 
linking number. This suggests that a knotted $(N_1,N_2)$ 
egoron exhibits internal anyonic exchange statistics.

Finally, we recall that the role of ghosts is to eliminate 
the unphysical gauge degrees of freedom from the Yang-Mills action. 
For a fully gauge fixed theory we need to complete (\ref{act1})
by two additional ghosts, which cancel the two unphysical
polarizations in the abelian $A^3_\mu$. The four ghosts 
displayed in (\ref{act1}) remove four unphysical field degrees 
of freedom among the eight which are
represented by the variables $\psi_{1,2}$ and $\bfe_\mu$. 
We propose that this entirely removes $\bfe_\mu$, leaving the
four $\psi_{1,2}$. From (\ref{dirac})
and (\ref{supc}) we then conclude that the Yang-Mills theory must
be intimately related to the abelian 
two-condensate superconductor with Lagrangian of the form
\begin{equation}
- \frac{1}{4} F_{\mu\nu}^2 + |(\partial_\mu + igA_\mu)\psi_1|^2
+ |(\partial_\mu - igA_\mu)\psi_2|^2 
- \lambda( |\psi_1|^2 - 
|\psi_2|^2)^2 +  \gamma(\psi^\star_1 \psi_2 + \psi_2^\star \psi_1)
\label{propo}
\end{equation}
This is indeed a manifestly unitary theory for
six physical field degrees of freedom, the two transverse 
components of the U(1) gauge field together with the four scalars. 
The $\lambda$ and $\gamma$ 
are determined dynamically, by the underlying Yang-Mills 
theory. Furthermore, if we compute the one-loop
Yang-Mills $\beta$-function from (\ref{propo}), despite it 
being an abelian theory we nevertheless 
obtain the correct SU(2) result provided we renormalize
(\ref{propo}) by respecting the underlying SU(2) gauge structure. This
can be confirmed either by a comparison with the results 
in \cite{pac}, or by a direct evaluation of the one-loop 
scalar field effective potential in (\ref{propo}) and then
scaling the fields in a manner which is consistent with 
the underlying SU(2) structure; as detailed in \cite{lisa} this
leads to the correct SU(2) $\beta$-function, further
supporting our proposal that (\ref{propo}) 
is the right superconductor model for SU(2) Yang-Mills theory.

\vskip 0.2cm
In conclusion, we have argued that SU(2) Yang-Mills theory
can be mapped into a two-gap superconductor with interband 
Josephson couplings. This model supports line vortices, 
which are then candidates for confining strings. The
vortices have also curious properties including {\it a 
priori} fractional flux, algebraic confinement, and the 
possibility that closed vortex rings exhibit exotic internal
statistics. Finally, we argued that the superconductor model
leads to the correct Yang-Mills $\beta$-function.
\vskip 0.2cm

We thank E. Babaev, T. Ekholm,  L. Faddeev and O. Viro
for discussion, and ICTP at Trieste for hospitality. 
This work is supported by grant VR-2003-3466.

\vfill\eject


\begin{thebibliography}{99} 

\bibitem{clay} Color confinement is one of the seven Millenium Problems;
see {\tt http://www.claymath.org}

\bibitem{conf} for recent review, see G. Ripka, {\it Dual 
Superconductor Models of Color Confinement}
(Lecture Notes in Physics 639, Springer Verlag 2004)

\bibitem{fad1} L.D. Faddeev  and A.J. Niemi, Phys. Lett. {\bf 
B525} (2002) 195; T.A. Bolokhov and 
L.D. Faddeev, Theor. Math. Phys. (to appear)

\bibitem{pac} H. Min, T. Lee and P.Y. Pac, Phys. Rev. {\bf D32} (1985) 440

\bibitem{val} F.V. Gubarev, L. Stodolski and V.I. Zakharov, Phys. Rev. Lett.
{\bf 86} (2001) 2220; L. Stodolsky, P. van Baal and V.I. Zakharov,
Phys. Lett. {\bf B552} (2003) 214  

\bibitem{dirac} P. Dirac, Phys. Rev. {bf 74} (1948) 817

\bibitem{egor1} E. Babaev, L. Faddeev and A.J. Niemi, Phys. Rev {\bf
B65} (2002) 100512

\bibitem{egor2} E. Babaev, Phys. Rev. Lett. {\bf 89} (2002) 067001

\bibitem{htc} F. Bouquet {\it et.al.} Phys. Rev. lett. {\bf 87} (2001) 047001;
Amy Y. Liu {\it et.al.} Phys. Rev. Lett. {\bf 87} 087005; P. Szabo {\it et.al.}
Phys. Rev. Lett. {\bf 87} (2001) 137005

\bibitem{legg} A.J. Leggett, Prog. Theor. Phys. {\bf 36} (1966) 901

\bibitem{lines} M. L\"ubcke, S.M. Nasir, A.J. Niemi and 
K. Torokoff, Phys. Lett. {\bf B534} (2002) 195; L. Faddeev and
A.J. Niemi, Phys. Rev. Lett. {\bf 85} (2000) 3416 

\bibitem{sami} A.J. Niemi, K. Palo and S. Virtanen, 
Phys. Rev. {\bf D61} (2000) 085020;  R.S. Ward,
Phys. Rev. {\bf D66} (2002) 041701 

\bibitem{nature} L.D. Faddeev and A.J. Niemi,
Nature {\bf 387} (1997) 58

\bibitem{eba} E. Babaev and A.J. Niemi, to appear

\bibitem{lerda} For a review, see 
A. Lerda, {\it Anyons, Quantum Mechanics of Particles
with Fractional Statistics} (Springer-Verlag, Berlin 1992)

\bibitem{viro} L. Faddeev, A.J. Niemi and O. Viro (unpublished)

\bibitem{lisa} L. Freyhult, Int. J. Mod. Phys. {\bf A17} (2002) 3681

\end{thebibliography}
\end{document}